\documentclass[12pt]{article}
\usepackage[english]{babel}
\usepackage{graphicx} 
\usepackage{amsmath,amssymb} 
\usepackage{epsfig} 
\usepackage{caption} 
\usepackage{float}  
\usepackage{placeins}
\title{Expanded calculations of pn-QRPA electron capture rates on
	$^{55}$Co for presupernova and supernova physics}
\author{Jameel-Un Nabi\thanks{Faculty of Engineering Sciences, GIK Institute of Engineering Sciences and Technology, Topi 23640, Swabi, NWFP, Pakistan. Email: jnabi00@gmail.com} \and Muhammad Sajjad}
\date{}

\begin{document}
	
	\maketitle
	
	\begin{abstract}
Due to its abundance and its relatively high capture rates,
$^{55}$Co is one of the key nuclide that can control the dynamics
of core collapse of a massive star. Previously we introduced our
microscopic calculations of capture rates on $^{55}$Co using the
proton-neutron quasi-particle random phase approximation (pn-QRPA)
theory. Here we present for the first time an expanded calculation
of the electron capture rates on $^{55}$Co on an extensive
temperature-density scale. These type of scale is appropriate for
interpolation purposes and of greater utility for simulation
codes.
\end{abstract}
\section{Introduction}
Supernova explosions are one of the most spectacular explosion in
nature when the luminosity of the star becomes comparable to that
of an entire galaxy (containing around $10^{11}$ stars). The
supernova outbursts last for very short span of time (several
months to a few years) and as such their chance of detection
becomes small (once in a few decades)[1]. Massive stars (of masses
above $10M_{\odot}$) develop an iron core towards the end of their
burning phases. The cores contract to the point that electrons
become a degenerate gas. The long lasting battle between
degeneracy pressure of the electrons and its self-gravity is
responsible for further evolution of these stars. When the mass of
the core exceeds the corresponding Chandrasekhar limit:
\begin{equation}
M_{Ch} = 1.45 (2 Y_{e}^{2})M_{\odot}
\end{equation}
(here $Y_{e}$ is the lepton-to-baryon ratio) the degenerate
electron pressure is incapable of opposing self-gravity and rapid
contraction of the core is ensued.

 Electron capture by free protons and by nuclei tends to decrease
 the degeneracy pressure and leads to the gravitational collapse
 of the core of massive stars triggering a supernova explosion.
 The capture of electrons also plays a key role in the
 neutronization of the core material, and also effects the
 formation of heavy elements beyond iron (including the so-called
 cosmochronometers which provide information about the age of the
 Galaxy and of the Universe) via the r-process at the final stage
 of supernova explosion.

The electron capture rates also determine the initial dynamics of
the collapse and also, via Eqn. (1), the size of collapsing core
and in turn the fate of the shock wave. The electron neutrinos,
produced as a result of capture reactions, escape and in turn
contribute to the cooling of the iron core. This lowers the
entropy of the stellar core and consequently favors the collapse
of the core. Bethe et al [2, 3] pointed out that stellar collapse
is very sensitive to the entropy of the core and lepton-to-baryon
ratio  . Fuller, Fowler, and Newman (FFN) [4] performed the
first-ever extensive calculation of stellar weak rates including
the capture rates, neutrino energy loss rates and decay rates for
a wide density and temperature domain. They made this detailed
calculations for 226 nuclei in the mass range $21 \leq A \leq 60$.
They also stressed the importance of the Gamow-Teller (GT) giant
resonance strength in the capture of the electron and estimated
the GT centroids using zeroth-order ($0\hbar\omega$ ) shell model.

At the final evolution of the massive star, the electron capture
is dominated by Fermi and GT transitions. The treatment of the
Fermi transitions, which are important in beta decays, are
straightforward while correct description of GT transitions poses
a challenging problem in nuclear structure. Later, Aufderheide et
al. [5] extended the FFN work for heavier nuclei with A $>$ 60.
They tabulated the 90 top electron capture nuclei averaged
throughout the stellar trajectory for $0.40 \leq Ye \leq 0.5$ (see
Table~25 therein). Since then theoretical efforts were
concentrated on the microscopic calculations of capture rates of
these iron-regime nuclide. Shell model (e.g. [6]) and the
proton-neutron quasiparticle random phase approximation theory
(pn-QRPA) (e.g. [7]) were used extensively and with relative
success for the microscopic calculation of stellar capture rates.

Nabi et al. [8] calculated weak interaction rates for 709 nuclei
with A = 18 to 100 in stellar matter using the pn-QRPA theory.
Since then these calculations were further refined with use of
more efficient algorithms, incorporation of latest experimental
data, and fine-tuning of model parameters.

$^{55}$Co is abundant in the presupernova conditions and is
thought to contribute effectively in the notorious dynamics of
core-collapse. Aufderheide and collaborators [5] placed $^{55}$Co
among the list of top ten  most important electron capture nuclei
during the presupernova evolution. Later Heger et al. [9] also
identified $^{55}$Co as the most important nuclide for electron
capture for massive stars ($25M_{\odot}$). Realizing the
importance of $^{55}$Co in astrophysical environments, Nabi,
Rahman and Sajjad [10] reported the calculation of electron and
positron capture rates on $^{55}$Co using the pn-QRPA theory.
However there was a need to perform an expanded calculation of
these important capture rates on a detailed temperature-density
grid suitable for collapse simulation codes. Due to the extreme
conditions prevailing in these scenarios, interpolation of
calculated rates within large intervals of temperature-density
points posed some uncertainty in the values of electron capture
rates for collapse simulators.

In this paper we present for the first time an expanded
calculation of electron capture rates on $^{55}$Co at suitable
intervals of temperature-density intervals. Section 2 deals with
the formalism of our calculation. In Section 3 we will be
presenting some of our results. We finally will be concluding in
Section 4 and at the end Table 3 presents our expanded calculation
of Fermi energies and electron capture rates on $^{55}$Co in
stellar matter.

\section{The quasi-particle random phase approximation with a separable interaction}
One of the main advantages of using the pn-QRPA theory is the
liberty of working the calculations in a huge model space of as
big as $7\hbar\omega$.  We considered a total of 30 excited states
in parent nucleus, corresponding to energies in excess of 10 MeV.
The inclusion of a very large model space of in our model provides
enough space to handle excited states in parent and daughter
nuclei (around 200). Transitions between these states play an
important role in the calculated weak rates. For the calculations
of these weak rates we made the following assumptions. \\
(i)  Only allowed GT and super allowed Fermi transitions were
calculated. The contribution from forbidden transitions was
assumed to be negligible. \\
(ii) At high enough temperature, the electron is no more bound to
nucleus and the atoms were dealt as being totally ionized with the
electrons obeying the Fermi-Dirac distribution. \\
(iii) The distortion of electron wave function due to Coulomb
interaction with a nucleus was represented by Fermi function in
the phase space integrals. \\
(iv) We incorporated experimental data for the excitation energies
and GT transitions whenever available to add on the reliability of
the calculations. The calculated excitation energies (along with
their $logft$ values) were replaced with the measured one when
they were within 0.5 MeV of each other. Missing measured states
were inserted. No theoretical levels were replaced (inserted)
beyond the level in experimental compilations for $^{55}$Co
without definite spin and parity assignment (2.98 MeV).\\
(v) Neutrinos and antineutrinos were assumed to escape freely from
the interior of star. Neutrino (antineutrino) capture was not
taken into account.

We used the quasi-particle proton-neutron random phase
approximation (pn-QRPA) with a separable interaction of the form
\begin{equation}
H^{QRPA} =H^{sp} +V^{pair} +V_{GT}^{ph} +V_{GT}^{pp},
\end{equation}
here $H^{sp}$ is the single-particle Hamiltonian, $V^{pair}$  is
the pairing force, $V_{GT}^{ph}$ is the particle-hole (ph)
Gamow-Teller force, and $V_{GT}^{pp}$  is the particle-particle
(pp) Gamow-Teller force. Single particle energies and wave
functions were calculated in the Nilsson model, which takes into
account nuclear deformations. Pairing was treated in the BCS
approximation. The proton-neutron residual interactions occurred
as particle-hole and particle-particle interaction. The
interactions were given separable form and were characterized by
two interaction constants $\chi$  and $\kappa$, respectively. The
selections of these two constants were done in an optimal fashion.
For details of the fine tuning of the Gamow-Teller strength
parameters, we refer to Ref. [11,12]. In this work, we took the
values of $\chi = 0.2 MeV$ and $\kappa = 0.07 MeV$. Other
parameters required for the calculation of weak rates are the
Nilsson potential parameters, the deformation, the pairing gaps,
and the Q-value of the reaction. Nilsson-potential parameters were
taken from Ref. [13] and the Nilsson oscillator constant was
chosen as $\hbar \omega=41A^{-1/3}(MeV)$ (the same for protons and
neutrons). The calculated half-lives depend only weakly on the
values of the pairing gaps [14]. Thus, the traditional choice of
$\Delta _{p} =\Delta _{n} =12/\sqrt{A} (MeV)$ was applied in the
present work. The deformation parameter for $^{55}$Co, $\delta$,
was taken to be 0.06, according to M\"{o}ller and Nix [15]. (See
also the discussion on choice of deformation parameter in Ref
[16].)

The RPA is formulated for excitations from the $J^{\pi}= 0$ ground
state of an even-even nucleus. When the parent nucleus has an odd
nucleon, the ground state can be expressed as a one-quasiparticle
(q.p.) state, in which the odd q.p. occupies the single-q.p. orbit
of the smallest energy. Then two types of transitions are
possible. One is the phonon excitations, in which the q.p. acts
merely as a spectator. The other is transitions of the q.p., where
phonon correlations to the q.p. transitions in first order
perturbation are introduced [17]. The phonon-correlated one-q.p.
states are defined by

\begin{subequations}
\begin{eqnarray}
&& |p_{c} \rangle \, =\, a_{p}^{+} {\left| - \right\rangle} \, +
\, \sum _{n,\omega }a_{n}^{+} A_{\omega }^{+} (\mu ){\left| -
\right\rangle}
\times  \nonumber \\
&& |\left\langle - \right| \left[a_{n}^{+} A_{\omega }^{+}
(\mu )\right]^{+} H_{31} a_{p}^{+} {\left| - \right\rangle} E_{p} (n,\omega ), \\
&& |n_{c} \rangle \, = \, a_{n}^{+} {\left| - \right\rangle} \,
+\, \sum _{p,\omega }a_{p}^{+} A_{\omega }^{+} (-\mu ){\left| -
\right\rangle}
\times  \nonumber \\
&& \left\langle - \right| \left[a_{p}^{+} A_{\omega }^{+} (-\mu
)\right]^{+} H_{31} a_{n}^{+} {\left| - \right\rangle} E_{n}
(p,\omega ) ,
\end{eqnarray}
\end{subequations}

\begin{equation}
E_{a} (b,\omega )\, =\, \frac{1}{(\, \varepsilon _{a} \,
-\,\varepsilon _{b} \, -\, \omega )}
\end{equation}
the first term of Eqn. (3) is a proton (neutron) q.p. state and
the second term represents correlations of RPA phonons admixed by
the phonon-q.p. coupling Hamiltonian $H_{31}$, which is obtained
from the separable ph and pp forces by the Bogoliubov
transformation [18]. The sums run over all phonons and neutron
(proton) q.p. states which satisfy $m_{p}-m_{n} = \mu$, where
$m_{p(n)}$ denotes the third component of the angular momentum and
$\pi_{p}.\pi_{n}= 1$. Derivations of the q.p. transitions
amplitudes for the correlated states are given in Ref.[18] for a
general force and a general mode of charge-changing transitions.
Also
\begin{equation}
{\left\langle n_{c}  \right|} t_{\pm } \sigma _{-\mu } {\left|
p_{c} \right\rangle} =(-1)^{\mu } {\left\langle p_{c}  \right|}
t_{\mp } \sigma _{\mu } {\left| n_{c}  \right\rangle}
\end{equation}
The excited states can be constructed as phonon-correlated
multi-quasiparticles states. The transition amplitudes between the
multi-quasiparticle states can then be reduced to those of
single-particle states. Low-lying states of an odd-proton
even-neutron nucleus ($^{55}Co$) can be constructed

(i) by exciting the odd proton from the ground state
(one-quasiparticle states),

(ii) by excitation of paired proton (three proton states), or,

(iii) by excitation of a paired neutron (one-proton two-neutron
states).

The multi-quasiparticle transitions can be reduced to ones
involving correlated (c) one-quasiparticle states:

\begin{eqnarray}
&&\left \langle p_{1}^{f} p_{2}^{f} n_{1c}^{f} \left|t_{\pm }
\sigma _{-\mu } \right|\right.
\left. p_{1}^{i} p_{2}^{i} p_{3c}^{i} \right\rangle = \nonumber \\
&& \delta (p_{1}^{f} ,p_{2}^{i} )\delta (p_{2}^{f},p_{3}^{i} )\,
\left\langle n_{1c}^{f} \left|t_{\pm } \sigma _{-\mu }
\right|\right.
\left. p_{1c}^{i} \right\rangle  \nonumber \\
&& -\delta (p_{1}^{f} ,p_{1}^{i} )\delta (p_{2}^{f} ,p_{3}^{i} )\,
\left\langle n_{1c}^{f} \left|t_{\pm } \sigma _{-\mu }
\right|\right. \left.
p_{2c}^{i} \right\rangle  \nonumber \\
&& +\delta (p_{1}^{f} ,p_{1}^{i} )\delta (p_{2}^{f} ,p_{2}^{i}
)\left\langle n_{1c}^{f} \left|t_{\pm } \sigma _{-\mu }
\right|\right. \left. p_{3c}^{i} \right\rangle
\end{eqnarray}

\begin{eqnarray}
&& {\left\langle p_{1}^{f} p_{2}^{f} n_{1c}^{f} \left|t_{\pm }
\sigma _{\mu } \right|\right. \left. p_{1}^{i} n_{1}^{i}
n_{2c}^{i}
 \right\rangle =}\nonumber \\
&& \delta (n_{1}^{f} ,n_{2}^{i} )\left[\delta (p_{1}^{f},
p_{1}^{i} )\left\langle p_{2c}^{f} \left|t_{\pm } \sigma _{\mu }
\right|\right. \left. n_{1c}^{i} \right\rangle \right. \nonumber  \\
&& {\left. \, \, \, -\delta (p_{2}^{f} ,p_{1}^{i} ) \left\langle
p_{1c}^{f} \left|t_{\pm } \sigma _{\mu } \right|\right. \left.
n_{1c}^{i} \right\rangle \right]-\delta (n_{1}^{f} ,n_{1}^{i} )}
\nonumber \\
&&{\times \left[\delta (p_{1}^{f} ,p_{1}^{i} )\left\langle
p_{2c}^{f} \left|t_{\pm } \sigma _{\mu } \right|\right. \left.
n_{2c}^{i}
\right\rangle \right. -\delta (p_{2}^{f} ,p_{1}^{i} )} \nonumber \\
&&{ \left. \times \left\langle p_{1c}^{f} \left|t_{\pm } \sigma
_{\mu } \right|\right. \left. n_{2c}^{i} \right\rangle \right]}
\end{eqnarray}

\begin{eqnarray}
&&{\left\langle n_{1}^{f} n_{2}^{f} n_{3c}^{f} \left|t_{\pm }
\sigma _{-\mu } \right|\right. \left. p_{1}^{i} n_{1}^{i}
n_{2c}^{i}
\right\rangle =} \nonumber \\
&&{ \delta (n_{2}^{f} ,n_{1}^{i} )\delta (n_{3}^{f} ,n_{2}^{i} )\,
\left\langle n_{1c}^{f} \left|t_{\pm } \sigma _{-\mu }
\right|\right.
\left. p_{1c}^{i} \right\rangle } \nonumber \\
&&{ -\delta (n_{1}^{f} ,n_{1}^{i} )\delta (n_{3}^{f} ,n_{2}^{i}
)\, \left\langle n_{2c}^{f} \left|t_{\pm } \sigma _{-\mu }
\right|\right.
\left. p_{1c}^{i} \right\rangle } \nonumber  \\
&&{+\delta (n_{1}^{f} ,n_{1}^{i} )\delta (n_{2}^{f} ,n_{2}^{i}
)\left\langle
 n_{3c}^{f} \left|t_{\pm } \sigma _{-\mu } \right|\right.
\left. p_{1c}^{i} \right\rangle .}
\end{eqnarray}

For the case of daughter nucleus ($^{55}Fe$), which is an
odd-neutron even-proton nucleus, the excited states can be
constructed

(i) by lifting the odd neutron from the ground state to excited
states (one  quasiparticle states),

(ii) by excitation of a paired neutron (three neutron states), or,

(iii) by the excitation of a paired proton (one-neutron two-proton
states).

Once again the multi-qp states are reduced to ones involving only
correlated (c) one-qp states:

\begin{eqnarray}
&& {\left\langle p_{1}^{f} n_{1}^{f} n_{2c}^{f} \left|t_{\pm }
\sigma _{\mu } \right|\right. \left. n_{1}^{i} n_{2}^{i}
n_{3c}^{i}
\right\rangle =} \nonumber \\
&& {  \delta (n_{1}^{f} ,n_{2}^{i} )\delta (n_{2}^{f} ,n_{3}^{i} )
\left\langle p_{1c}^{f} \left|t_{\pm } \sigma _{\mu }
\right|\right.
\left. n_{1c}^{i} \right\rangle } \nonumber \\
&& { -\delta (n_{1}^{f} ,n_{1}^{i} )\delta (n_{2}^{f} ,n_{3}^{i} )
\left\langle p_{1c}^{f} \left|t_{\pm } \sigma _{\mu }
\right|\right.
\left. n_{2c}^{i} \right\rangle } \nonumber \\
&& { +\delta (n_{1}^{f} ,n_{1}^{i} )\delta (n_{2}^{f} ,n_{2}^{i}
)\left \langle p_{1c}^{f} \left|t_{\pm } \sigma _{\mu }
\right|\right. \left. n_{3c}^{i} \right\rangle }
\end{eqnarray}

\begin{eqnarray}
&&{\left\langle p_{1}^{f} n_{1}^{f} n_{2c}^{f} \left|t_{\pm }
\sigma _{-\mu }
\right|\right. \left. p_{1}^{i} p_{2}^{i} n_{1c}^{i} \right\rangle =}\nonumber \\
&&{ \delta (p_{1}^{f} ,p_{2}^{i} )\left[\delta (n_{1}^{f}
,n_{1}^{i} )\right.
 \left\langle n_{2c}^{f} \left|t_{\pm } \sigma _{-\mu } \right|\right.
\left. p_{1c}^{i} \right\rangle } \nonumber \\
&&{ -\delta (n_{2}^{f} ,n_{1}^{i} )\left\langle n_{1c}^{f}
\left|t_{\pm } \sigma _{-\mu } \right|\right. \left. p_{1c}^{i}
\right\rangle
-\delta (p_{1}^{f} ,p_{1}^{i} )} \nonumber \\
&&{ \times \left[\delta (n_{1}^{f} ,n_{1}^{i} )\right.
\left\langle n_{2c}^{f} \left|t_{\pm } \sigma _{-\mu }
\right|\right. \left.
p_{2c}^{i} \right\rangle -\delta (n_{2}^{f} ,n_{1}^{i} )} \nonumber \\
&&{ \left. \times \left\langle n_{1c}^{f} \left|t_{\pm } \sigma
_{-\mu } \right|\right. \left. p_{2c}^{i} \right\rangle \right].}
\end{eqnarray}

\begin{eqnarray}
&& {\left\langle p_{1}^{f} p_{2}^{f} p_{3c}^{f} \left|t_{\pm }
\sigma _{\mu }
 \right|\right. \left. p_{1}^{i} p_{2}^{i} n_{1c}^{i} \right\rangle =}
\nonumber \\
&&{ \delta (p_{2}^{f} ,p_{1}^{i} )\delta (p_{3}^{f} ,p_{2}^{i} )
\left\langle p_{1c}^{f} \left|t_{\pm } \sigma _{\mu }
\right|\right.
\left. n_{1c}^{i} \right\rangle } \nonumber \\
&&{ -\delta (p_{1}^{f} ,p_{1}^{i} )\delta (p_{3}^{f} ,p_{2}^{i} )
\left\langle p_{2c}^{f} \left|t_{\pm } \sigma _{\mu }
\right|\right. \left. n_{1c}^{i} \right\rangle } \nonumber \\
&&{ +\delta (p_{1}^{f} ,p_{1}^{i} )\delta (p_{2}^{f} ,p_{2}^{i} )
\left\langle p_{3c}^{f} \left|t_{\pm } \sigma _{\mu }
\right|\right. \left. n_{1c}^{i} \right\rangle .}
\end{eqnarray}

For all the given q.p. transition amplitudes [Eqs.~(6)-~(11)],
the antisymmetrization of the single- q.p. states was taken into account:\\
$ p_{1}^{f}<p_{2}^{f}<p_{3}^{f}<p_{4}^{f}$,\\
$ n_{1}^{f}<n_{2}^{f}<n_{3}^{f}<n_{4}^{f}$,\\
$ p_{1}^{i}<p_{2}^{i}<p_{3}^{i}<p_{4}^{i}$,\\
$ n_{1}^{i}<n_{2}^{i}<n_{3}^{i}<n_{4}^{i}$.\\
GT transitions of phonon excitations for every excited state were
also taken into account. We also assumed that the quasiparticles
in the parent nucleus remained in the same quasiparticle orbits.

The capture rates of a transition from the $ith$ state of the
parent to the $jth$ state of the daughter nucleus is given by
\begin{equation}
\lambda ^{^{ec} } _{ij} =\left[\frac{\ln 2}{D} \right]\left[f_{ij}
(T,\rho ,E_{f} )\right]\left[B(F)_{ij} +\left({\raise0.7ex\hbox{$
g_{A}  $}\!\mathord{\left/ {\vphantom {g_{A}  g_{V} }} \right.
\kern-\nulldelimiterspace}\!\lower0.7ex\hbox{$ g_{V}  $}}
\right)^{2} B(GT)_{ij} \right].
\end{equation}
We took the value of D=6295s [19] and the ratio of the axial
vector to the vector coupling constant as -1.254 [20]. $B_{ij}'s$
are the sum of reduced transition probabilities of the Fermi B(F)
and GT transitions B(GT). Details of these reduced transition
probabilities can be found in Ref. [7, 21]. The phase space
integral $f_{ij}$ is an integral over total energy and for
electron capture it is given by
\begin{equation}
 f_{ij} \, =\, \int _{w_{l} }^{\infty }w\sqrt{w^{2} -1}
 (w_{m} \, +\, w)^{2} F(+Z,w)G_{-} dw.
\end{equation}
In above equation $w$ is the total energy of the electron
including its rest mass, $w_{l}$ is the total capture threshold
energy (rest+kinetic) for electron capture. F(+Z,w) are the Fermi
functions and were calculated according to the procedure adopted
by Gove and Martin [22]. G$_{-}$ is the Fermi-Dirac distribution
function for electrons,
\begin{equation}
 G_{-} =\left[\exp \left(\frac{E-E_{f} }{kT}
 \right)+1\right]^{-1}.
\end{equation}
Here $E = (w-1)$ is the kinetic energy of the electrons, $E_{f}$
is the Fermi energy of the electrons, $T$ is the temperature, and
$k$ is the Boltzmann constant.

The number density of electrons associated with protons and nuclei
is $\rho Y_{e}N_{A}$ ($\rho$ is the baryon density and $N_{A}$ is
Avogadros number) where,
\begin{equation}
\rho Y_{e}=\frac{1}{\pi ^{2} N_{A} }(\frac{m_{e}c}{\hbar})^{3}\int
_{0}^{\infty }(G_{-}  -G_{+} )p^{2} dp,
\end{equation}
here $p = (w^{2}-1)^{1/2}$ is the electron momentum and Eqn. (15)
has the units of $mol \hspace{0.1in} cm^{-3}$. G$_{+}$ is the
Fermi-Dirac distribution function for positrons,

\begin{equation}
G_{+} =\left[\exp \left(\frac{E+2+E_{f} }{kT}
\right)+1\right]^{-1}.
\end{equation}
Eqn. (15) is used for an iterative calculation of Fermi energies
for selected values of $Y_{e}$ and T. There is a finite
probability of occupation of parent excited states in the stellar
environment as result of the high temperature in the interior of
massive stars. Capture rates then also have a finite contribution
from these excited states. The total electron capture rate per
unit time per nucleus is given by
\begin{equation}
\lambda_{ec} =\sum _{ij}P_{i} \lambda _{ij}.
\end{equation}
The summation over all initial and final states was carried out
until satisfactory convergence in the rate calculations was
achieved. Here $P_{i}$ is the probability of occupation of parent
excited states and follows the normal Boltzmann distribution.

\section{Results and discussions}
The luxury of having a huge model space at our disposal allowed us
to perform the calculation of electron capture rates from 30
excited states of $^{55}$Co. At temperatures pertinent to
supernova environment we have a finite probability of occupation
of excited states and a microscopic calculation of rates from
these excited states is desirable. Earlier, Nabi and Sajjad [16]
did point to the fact that the Brink's hypothesis ( and back
resonances for calculation of beta decay rates) usually employed
in large-scale shell model calculations is not a good
approximation to use in stellar rate calculations. Brink's
hypothesis states that GT strength distribution on excited states
is identical to that from ground state, shifted only by the
excitation energy of the state whereas the GT back resonances are
states reached by the strong GT transitions in the electron
capture process built on ground and excited states. Table 1 shows
the calculated excited states of $^{55}$Co using the pn-QRPA
theory. For each parent state we considered around 200 excited
states in daughter $^{55}$Fe.

The GT strength distribution from ground state and first two
excited states of $^{55}$Co was presented earlier (see Fig.1 of
Ref [10]). We calculated the total GT strength from ground state
to be 7.4 with a centroid in the energy range 7.1 - 7.4 MeV in
daughter $^{55}$Fe. Table 2 shows the calculated B(GT) strength
values for the ground state of $^{55}$Co. The strengths are given
up to energy of 10 MeV in daughter nucleus, $^{55}$Fe. Calculated
B(GT) strength of magnitude less than $10^{-3}$ are not included
in this table.

Comparison of our calculation of electron capture rates with the
pioneer work of FFN [4] and large-scale shell model calculations
[23] was also presented in detail in Ref [10]. The pn-QRPA rates
were enhanced up to two orders of magnitude compared to the
large-scale shell model calculations. Our rates were also enhanced
as compared to the FFN calculations. However at high temperatures
and densities( logT $>$ 9.4; $\rho \approx 10^{11} gcm^{-3}$) the
FFN rates surpass our rates. The main reason for this enhancement
was that FFN placed the GT centroid at too low excitation energies
[10, 23].

Fig.1 shows four panels depicting our calculated electron capture
rates at selected temperature and density domain. The upper left
panel shows the electron capture rates in low-density region
($\rho [gcm^{-3}] =10^{0.5}, 10^{1.5}$ and $10^{2.5}$), the upper
right in medium-low density region ($\rho [gcm^{-3}] =10^{3.5},
10^{4.5}$ and $10^{5.5}$), the lower left in medium-high density
region ($\rho [gcm^{-3}] =10^{6.5}, 10^{7.5}$ and $10^{8.5}$) and
finally the lower right panel depicts our calculated electron
capture rates in high density region ($\rho [gcm^{-3}] =10^{9.5},
10^{10.5}$ and $10^{11}$). It can be seen from this figure that in
the low density region the capture rates, as a function of
temperature, are more or less superimposed on one another. This
means that there is no appreciable change in the rates with
increasing the density by an order of magnitude. However as we go
from the medium low density region to high density region these
rates start to 'peel off' from one another. Orders of magnitude
difference in rates are observed (as a function of density) in
high density regions. For a given density the rates increase
monotonically with increasing temperatures. We also notice that
the rates coincide at T$_{9}$ = 30K except for the high density
region (where T$_{9}$ gives the stellar temperature in units of
$10^{9}$K).

Fig.2 shows our summed GT strength as a function of excited states
in the daughter $^{55}$Fe. We note that almost all the strength
cumulates up to an energy around 12 MeV in $^{55}$Fe. No
appreciable strength is seen in $^{55}$Fe at higher excitation
energies.

We finally present our calculated electron capture rates on a
detailed temperature-density grid in Table 3. Here Column 1 shows
the density in logarithmic scales (in units of $gcm^{-3}$), Column
2, the stellar temperature in units of $10^{9}K$, Column 3 gives
the calculated Fermi energy at the corresponding temperature and
density whereas Column 4 displays the calculated electron capture
rates in logarithmic scales (in units of $sec^{-1}$). Tables of
rate calculations presented earlier in compilations (e.g. [24, 7,
4]) were not presented on a detail temperature-density grid and at
times lead to erroneous results when interpolated. We hope that
this table will prove more useful for core-collapse simulators.

\section{Conclusions}
$^{55}$Co is advocated to play a key role amongst the iron-regime
nuclide controlling the dynamics of core-collapse of massive
stars. The electron capture rates on $^{55}$Co are used as nuclear
physics input parameter for core-collapse simulations. Reliable
and detailed calculation of electron capture rates is desirable
for these codes. These parameters may fine tune the final outcome
of these simulations.

Here we present, for the first time, an expanded calculation of
stellar electron capture rates on a fine temperature-density scale
suitable for simulation codes. We considered a total of 30 parent
excited states for the microscopic calculation of these stellar
capture rates on $^{55}$Co. Our calculations point towards an
enhanced capture rate for $^{55}$Co as compared to large-scale
shell model calculations. This may have a significant impact on
the outcome of simulation results. We remind the readers that it
was the rather reduced electron capture rates on $^{55}$Co (from
previous calculations) in the outer layers of the core of massive
stars that lead to slowing of the collapse and resulted,
consequently, in a large shock radius to deal with [25]. Can our
(relatively enhanced) electron capture change this scenario? It
might. We again recall that according to the calculations of
Aufderheide and collaborators [5], the rate of change of lepton
fraction can change by as much as 50 $\%$ alone due to electron
capture rates on $^{55}$Co. We will urge core-collapse simulators
to test run our reported electron capture rates presented here to
check for some interesting outcome.

\newpage
\textbf{Table 1:} Calculated excited states in parent $^{55}Co$.
\begin{center}
\begin{tabular}{ccccccccccccccccccc} \\ \hline
0.0  &  & 2.17& & 2.57 & & 2.92 & &   3.08 & &  3.30 & & 3.87 & &
4.10& & 4.48 & &4.89\\
5.20  & & 5.47 & & 5.68  & & 5.99 & & 6.20 & & 6.50 & & 6.79 & &
7.08 & & 7.33 & &   7.65\\
7.88 & & 8.16 & &   8.42 & & 8.67 & & 8.96 & &  9.21 & & 9.48 & &
9.74& & 9.89 & & 10.00\\ \hline
\end{tabular}
\end{center}
\vspace{1.5in} \textbf{Table 2:} Calculated B(GT+) values from
ground state in $^{55}Co$.
\begin{center}
\begin{tabular}{cc|cc|cc} \\ \hline
Energy(MeV)& B(GT+) & Energy(MeV)& B(GT+) &Energy(MeV)& B(GT+) \\ \hline
2.46&    2.35E-01&    6.36&   5.45E-01&    8.25&    1.06E-01\\
3.01&    2.13E-01&    6.66&   3.22E-01&    8.36&    2.06E-03\\
3.45&    3.05E-01&    6.85&   1.70E-01&    8.62&    8.83E-03\\
3.57&    4.50E-03&    7.00&   2.01E-03&    8.73&    2.58E-02\\
3.68&    1.84E-01&    7.14&   8.70E-01&    8.90&    2.45E-02\\
3.90&    2.52E-01&    7.43&   7.07E-01&    9.11&    1.38E-02\\
4.11&    2.50E-02&    7.59&   1.87E-02&    9.26&    3.04E-02\\
5.82&    2.94E-01&    7.76&   3.89E-01&    9.42&    1.52E-02\\
5.93&    2.60E-03&    7.92&   1.49E-02&    9.60&    1.70E-02\\
6.13&    1.98E-01&    8.04&   1.69E-01                      \\
\hline

\end{tabular}
\end{center}
 \textbf{Table 3:} Calculated electron rates on
$^{55}Co$ for different selected densities and temperatures in
stellar matter. ADen is the log($\rho Y_{e}$) and has units of $g
cm^{-3}$, where $\rho$ is the baryon density and $Y_{e}$ is the
ratio of the electron number to the baryon number. Temperatures
($T_{9}$) are measured in $10^{9}$ K. EFermi is the total Fermi
energy of electron including the rest mass (MeV). E-Cap  are the
electron capture rates. The calculated electron capture rates are
tabulated in $log_{10}\lambda _{ec}$ in units of $sec^{-1}$.\\
\begin{table}[H]
	\centering
	\begin{tabular}{cccc|cccc|cccc}
		\hline
		ADen & $T_{9}$ & EFermi & E-Cap  & ADen & $T_{9}$ & EFermi & E-Cap   & ADen & $T_{9}$ & EFermi & E-Cap  \\ \hline
     0.5  & 0.5     & 0.065  & -8.726  &1    &8.5      &  0     &  -0.97   &2    & 4.5     & 0       &-2.44    \\
     0.5  & 1       &  0     & -6.321  &1    &9        &  0     &  -0.824  &2    & 5       & 0       &-2.206   \\
     0.5  & 1.5     &  0     & -5.068  &1   & 9.5     & 0       & -0.681   &2   & 5.5     & 0       &-1.994   \\
     0.5  & 2       &  0     & -4.308  &1   & 10      & 0       & -0.541   &2   & 6      &    0    &-1.798  \\
     0.5  & 2.5     &  0     & -3.766  &1   &  15     & 0       &  0.683   &2   &  6.5    &   0    & -1.615    \\
     0.5  & 3       &  0     & -3.344  &1   &  20     & 0       &  1.601   &2   &   7     &   0    & -1.443  \\
     0.5  & 3.5     &  0     & -2.997  &1   &  25     & 0       &  2.284   &2   &   7.5   &   0    & -1.279 \\
     0.5  & 4       &  0     & -2.701  &1   & 30      & 0       &  2.813   &2   &   8     &   0    & -1.122   \\
     0.5  & 4.5     &  0     & -2.44   &1.5 & 0.5     & 0.162   & -7.747   &2   &   8.5   &   0    & -0.97\\
     0.5  & 5       &  0     & -2.207  &1.5  & 1       & 0.002   & -6.314  &2   &   9     &   0    & -0.823  \\
     0.5  & 5.5     &  0     & -1.994  &1.5  & 1.5     & 0       & -5.068  &2   &   9.5   &   0    & -0.68 \\
     0.5  & 6       &  0     & -1.798  &1.5  & 2       & 0       & -4.307  &2   &   10    &   0    & -0.541 \\
     0.5  & 6.5     &  0     & -1.616  &1.5  & 2.5     & 0       & -3.765  &2   &   15    &   0    &  0.683 \\
     0.5  & 7       &  0     & -1.443  &1.5  & 3       & 0       & -3.344  &2   &   20    &   0    &  1.601 \\
     0.5  & 7.5     &  0     & -1.279  &1.5  & 3.5     & 0       & -2.997  &2   &   25    &   0    &  2.285 \\
     0.5  & 8       &  0     & -1.122  &1.5  & 4       & 0       & -2.7    &2   &   30    &   0    &  2.813 \\
     0.5  & 8.5     &  0     & -0.971  &1.5  & 4.5     & 0       & -2.44   &2.5 &  0.5    &   0.261& -6.748 \\
     0.5  & 9       &  0     & -0.824  &1.5  & 5       & 0       & -2.206  &2.5 &  1      &   0.015& -6.246\\
     0.5  & 9.5     &  0     & -0.681  &1.5  & 5.5     & 0       & -1.994  &2.5 &  1.5    &   0.002& -5.064\\
     0.5  &10       &  0     & -0.542  &1.5  & 6       & 0       &-1.798   &2.5 &  2      &   0    & -4.306 \\
     0.5  &15       &  0     &  0.682  &1.5  & 6.5     & 0       & -1.615  &2.5 &  2.5    &   0    & -3.765\\
     0.5  &20       &  0     &  1.6    &1.5  & 7       & 0       & -1.443  &2.5 &  3      &   0    & -3.344\\
     0.5  &25       &  0     &  2.283  &1.5  & 7.5     & 0       & -1.279  &2.5  &  3.5    &   0    & -2.997\\
     0.5  &30       &  0     &  2.812  &1.5  & 8       & 0       & -1.122  &2.5  &  4      &   0    & -2.7\\
     1    &0.5      &  0.113 &  -8.245 &1.5  & 8.5     & 0       & -0.97   &2.5  &  4.5    &   0    & -2.439\\
     1    &1        &  0     &  -6.32  &1.5  & 9       & 0       & -0.823  &2.5  &  5      &   0    & -2.206\\
     1    &1.5      &  0     &  -5.069 &1.5  & 9.5     & 0       &-0.68    &2.5  &  5.5    &   0    & -1.994\\
     1    &2        &  0     &  -4.307 &1.5  & 10      & 0       &-0.541   &2.5  &  6      &   0    & -1.798\\
     1    &2.5      &  0     &  -3.765 &1.5  & 15      & 0       &0.683    &2.5  &  6.5    &   0    & -1.615\\
     1    &3        &  0     &  -3.344 &1.5  & 20      & 0       &1.601    &2.5  &  7      &   0    & -1.443\\
     1    &3.5      &  0     &  -2.997 &1.5  & 25      & 0       &2.285    &2.5  &  7.5    &   0    & -1.279\\
     1    &4        &  0     &  -2.7   &1.5  & 30      & 0       &2.813    &2.5  &  8      &   0    & -1.122\\
     1    &4.5      &  0     &  -2.44  &2    & 0.5     & 0.212   &-7.248   &2.5  &  8.5    &   0    & -0.97\\
     1    &5        &  0     &  -2.206 &2    & 1       & 0.005   &-6.298   &2.5  &  9      &   0    & -0.823\\
     1    &5.5      &  0     &  -1.994 &2    & 1.5     & 0       &-5.067   &2.5  &  9.5    &   0    & -0.68\\
     1    &6        &  0     &  -1.798 &2    & 2       & 0       &-4.307   &2.5  &  10     &   0    & -0.541\\
     1    &6.5      &  0     &  -1.615 &2    & 2.5     & 0       &-3.765   &2.5  &  15     &   0    &  0.683\\
     1    &7        &  0     &  -1.443 &2    & 3       & 0       &-3.344   &2.5  &  20     &   0    &  1.601\\
     1    &7.5      &  0     &  -1.279 &2    & 3.5     & 0       &-2.997   &2.5  &  25     &   0    &  2.285\\
     1    &8        &  0     &  -1.122 &2    & 4       & 0       &-2.7     &2.5  &  30     &   0    &  2.813\\

\end{tabular}
\end{table}
\begin{table}[H]
	\centering

\begin{tabular}{cccc|cccc|cccc} \\ \hline
ADen & $T_{9}$ & EFermi & E-Cap  & ADen & $T_{9}$ & EFermi & E-Cap & ADen & $T_{9}$ & EFermi & E-Cap  \\ \hline
3&  0.5&  0.311  &  -6.248 &    4 &   0.5&  0.411&    -5.252&   5&    0.5&  0.522&    -4.281\\
3&    1&    0.046&    -6.092&   4 &   1  &  0.209 &   -5.274 &  5&    1 &   0.413&    -4.282\\
3&    1.5&  0.005&    -5.053&   4 &   1.5&  0.047 &   -4.912&   5&    1.5&  0.265&    -4.193\\
3&    2&    0.001&    -4.304&   4 &   2  &  0.014 &   -4.273&   5&    2 &   0.128&    -3.989\\
3&    2.5&  0.001&    -3.764&   4 &   2.5&  0.006 &   -3.753&   5&    2.5&  0.062&    -3.643\\
3&    3 &   0  &  -3.343  & 4 &   3  &  0.004&    -3.338&   5&    3 &   0.035 &   -3.286\\
3&    3.5&  0  &  -2.996 &  4  &  3.5&  0.002&    -2.994&   5&    3.5&  0.023&    -2.965\\
3&    4  &  0  &  -2.7    & 4  &  4  &  0.002&    -2.698 &  5&    4 &   0.016 &   -2.68\\
3&    4.5&  0  &  -2.439 &  4  &  4.5&  0.001&    -2.438 &  5 &   4.5&  0.012&    -2.427\\
3&    5  &  0  &  -2.206  & 4  &  5  &  0.001&    -2.205 &  5&    5  &  0.009&    -2.197\\
3&    5.5&  0  &  -1.994  & 4  &  5.5&  0.001&    -1.993 &  5&    5.5&  0.007&    -1.987\\
3&    6  &  0  &  -1.798  & 4  &  6   & 0.001&    -1.797 &  5&    6  &  0.006&    -1.793\\
3&    6.5 & 0  &  -1.615 &  4  &  6.5 & 0.001&    -1.614 &  5 &   6.5&  0.005&    -1.611\\
3&    7  &  0  &  -1.443  & 4  &  7  &  0 &   -1.442&   5 &   7 &   0.004&    -1.44\\
3&    7.5&  0  &  -1.279  & 4  &  7.5 & 0 &   -1.278 &  5&    7.5&  0.004&    -1.276\\
3&    8  &  0  &  -1.122  & 4  &  8   & 0 &   -1.121 &  5&    8  &  0.003&    -1.12\\
3&    8.5&  0  &  -0.97  &  4  &  8.5&  0 &   -0.97  &  5&    8.5&  0.003&    -0.969\\
3&    9  &  0  &  -0.823 &  4  &  9   & 0  &  -0.823 &  5&    9  &  0.002&    -0.822\\
3&    9.5&  0  &  -0.68   & 4  &  9.5 & 0  &  -0.68  &  5&    9.5&  0.002&    -0.679\\
3&    10 &  0  &  -0.541  & 4  &  10 &  0  &  -0.541 &  5&    10 &  0.002&    -0.54\\
3&    15 &  0  &  0.683   & 4  &  15  & 0  &  0.683  &  5&    15&   0.001&     0.683\\
3&    20  & 0  &  1.601   & 4  &  20  & 0  &  1.601  &  5&    20&   0 &   1.601\\
3&    25 &  0  &  2.285  &  4 &   25  & 0  &  2.285  &  5&    25 &  0 &   2.285\\
3&    30 &  0  &  2.813  &  4 &   30  & 0  &  2.813  &  5&    30 & 0 &   2.813\\

3.5&  0.5&  0.361&    -5.749&   4.5&  0.5&  0.464&    -4.76 &   5.5&  0.5&  0.598&    -3.819\\
3.5&  1  &  0.115&    -5.745&   4.5&  1 &   0.309&    -4.778&   5.5&  1 &   0.528&    -3.79\\
3.5&  1.5&  0.015 &   -5.018&   4.5&  1.5&  0.13 &    -4.637&   5.5&  1.5&  0.423&    -3.7\\
3.5&  2  &  0.004&    -4.296&   4.5&  2  &  0.044&    -4.199&   5.5&  2  &  0.295&    -3.585\\
3.5&  2.5&  0.002&    -3.761&   4.5&  2.5&  0.02 &    -3.726&   5.5&  2.5&  0.18 &    -3.412\\
3.5&  3  &  0.001 &   -3.342&   4.5&  3  &  0.011 &   -3.325 &  5.5&  3  &  0.11 &    -3.165\\
3.5&  3.5&  0.001 &   -2.996&   4.5&  3.5&  0.007 &   -2.987&   5.5&  3.5&  0.072&    -2.896\\
3.5&  4  &  0.001 &   -2.699 &  4.5 & 4   & 0.005 &   -2.694 &  5.5&  4  &  0.05&     -2.638\\
3.5&  4.5&  0   & -2.439 &  4.5&  4.5&  0.004 &   -2.435&   5.5&  4.5&  0.038&    -2.399\\
3.5&  5  &  0   & -2.206 &  4.5&  5 &   0.003 &   -2.203&   5.5&  5  &  0.029 &   -2.178\\
3.5&  5.5&  0   & -1.993 &  4.5&  5.5&  0.002 &   -1.992&   5.5&  5.5&  0.023&    -1.973\\
3.5&  6  &  0   & -1.797 &  4.5&  6  &  0.002 &   -1.796&   5.5&  6  &  0.019&    -1.782\\
3.5&  6.5&  0   & -1.615 &  4.5&  6.5&  0.002 &   -1.614 &  5.5&  6.5&  0.016&    -1.603\\
3.5&  7  &  0   & -1.442 &  4.5&  7  &  0.001 &    -1.442 &  5.5&  7  &  0.013&    -1.433\\
3.5&  7.5&  0   & -1.279 &  4.5&  7.5 & 0.001  &   -1.278&   5.5&  7.5&  0.012&    -1.271\\
3.5&  8  &  0   & -1.122 &  4.5&  8  &  0.001  &   -1.121 &  5.5&  8  &  0.01 &    -1.116\\
3.5&  8.5&  0   & -0.97  &  4.5&  8.5 & 0.001  &   -0.97 &   5.5&  8.5&  0.009 &   -0.965\\
3.5&  9  &  0   & -0.823 &  4.5&  9  &  0.001  &   -0.823&   5.5&  9  &  0.008 &   -0.819\\
3.5&  9.5&  0   & -0.68  &  4.5&  9.5 & 0.001  &   -0.68 &   5.5&  9.5&  0.007 &   -0.677\\
3.5&  10 &  0   & -0.541 &  4.5&  10 &  0.001  &   -0.541&   5.5&  10 &  0.006&    -0.538\\
3.5&  15 &  0   & 0.683 &   4.5&  15 &  0 &   0.683 &   5.5&  15&   0.003&    0.684\\
3.5&  20 &  0   & 1.601  &  4.5&  20 &  0  &  1.601 &   5.5&  20&   0.001&     1.602\\
3.5&  25 &  0   & 2.285  &  4.5&  25 &  0  &  2.285  &  5.5&  25 &  0.001 &    2.285\\
3.5&  30 &  0   & 2.813  &  4.5 & 30 &  0  &  2.813  &  5.5&  30& 0.001 & 2.813\\
\end{tabular}
\end{table}
\begin{table}[H]
	\centering
\begin{tabular}{cccc|cccc|cccc} \\ \hline
ADen & $T_{9}$ & EFermi & E-Cap  & ADen & $T_{9}$ & EFermi & E-Cap & ADen & $T_{9}$ & EFermi & E-Cap  \\ \hline

6&   0.5&  0.713&    -3.352&   7 &   0.5&  1.217&    -2.228&   8&    0.5&  2.444&    -0.744\\
6&    1 &   0.672&    -3.297&   7&    1 &   1.2 &     -2.172&   8&    1 &   2.437&    -0.709\\
6&    1.5&  0.604&    -3.195&   7&    1.5&  1.173&    -2.076&   8&    1.5&  2.424&    -0.646\\
6&    2  &  0.512&    -3.091&   7&    2  &  1.133 &   -1.982&   8&    2  &  2.406&    -0.587\\
6&    2.5&  0.405&    -2.987&   7&    2.5&  1.083&    -1.901&   8&    2.5&  2.383&   -0.538\\
6&    3  &  0.299&    -2.861&   7&    3  &  1.021&    -1.827&   8&    3  &  2.355&    -0.495\\
6&    3.5&  0.214&    -2.699&  7 &   3.5&  0.95  &   -1.757 &  8&    3.5&  2.322 &   -0.457\\
6&    4  &  0.156&   -2.51  &  7 &   4  &  0.871 &   -1.688 &  8&    4  &  2.283 &   -0.419\\
6&    4.5&  0.117&    -2.312&   7&    4.5&  0.785&    -1.617&   8&    4.5&  2.24 &    -0.382\\
6&    5  &  0.091&    -2.117&   7&    5  &  0.698&    -1.542 &  8&    5  &  2.192&    -0.344\\
6&    5.5&  0.073&    -1.929&   7 &   5.5&  0.613&    -1.46  &  8&    5.5&  2.139&    -0.304\\
6&    6  &  0.06 &    -1.749&   7&    6  &  0.534&    -1.369 &  8&    6  &  2.081&    -0.263\\
6&    6.5&  0.05 &    -1.577&   7 &   6.5&  0.465&   -1.269 &  8&    6.5&  2.019 &   -0.22\\
6&    7  &  0.042&    -1.413&   7 &   7  &  0.404 &   -1.163&   8&    7  &  1.952 &   -0.174\\
6&    7.5&  0.036&    -1.255&   7 &   7.5&  0.353 &   -1.05 &   8&    7.5&  1.882 &   -0.126\\
6&    8  &  0.032&    -1.102&   7 &   8  &  0.31 &    -0.933 &  8&    8  &  1.808&    -0.074\\
6&    8.5&  0.028&    -0.954&   7 &   8.5&  0.274 &   -0.813 &  8&    8.5&  1.732&    -0.019\\
6&    9  &  0.025&    -0.81 &   7&    9  &  0.244 &   -0.691 &  8&    9  &  1.653&    0.039\\
6&    9.5&  0.022&    -0.669&   7 &   9.5&  0.218 &   -0.569 &  8&    9.5&  1.574&    0.101\\
6&    10 &  0.02 &    -0.531&   7 &   10 &  0.196 &   -0.445 &  8&    10 &  1.493&    0.166\\
6&    15 &  0.009&    0.686 &   7 &   15 &  0.085 &   0.711  &  8&    15 &  0.817&    0.946\\
6&    20 &  0.005&    1.603 &   7 &   20 &  0.047 &   1.613  &  8&    20 &  0.47 &   1.716\\
6&    25 &  0.003&    2.285 &   7 &   25 &  0.03 &    2.291  &  8&    25 &  0.301&    2.344\\
6&    30 &  0.002&    2.813 &   7 &   30 &  0.021&    2.816  &  8&    30 &  0.209&    2.847\\

6.5&  0.5&  0.905&    -2.834&   7.5&  0.5&  1.705&    -1.524 &  8.5&  0.5&  3.547&    0.075\\
6.5&  1  &  0.88 &    -2.772&   7.5&  1  &  1.693&    -1.479 &  8.5&  1  &  3.542&    0.103\\
6.5&  1.5&  0.837&    -2.667&   7.5&  1.5&  1.675&    -1.398 &  8.5&  1.5&  3.534&    0.154\\
6.5&  2  &  0.777&    -2.564&   7.5&  2  &  1.648&    -1.322 &  8.5&  2  &  3.521&    0.202\\
6.5&  2.5&  0.701&    -2.471&   7.5&  2.5&  1.614&    -1.257 &  8.5&  2.5&  3.506&    0.24\\
6.5&  3  &  0.612&    -2.384&   7.5&  3  &  1.573&    -1.199 &  8.5&  3  &  3.487&    0.271\\
6.5&  3.5&  0.517&    -2.293&   7.5&  3.5 & 1.524&    -1.145 &  8.5&  3.5&  3.464&   0.299\\
6.5&  4  &  0.424&    -2.191&   7.5&  4  &  1.468 &   -1.093 &  8.5&  4  &  3.438&    0.326\\
6.5&  4.5&  0.343&    -2.071&   7.5&  4.5&  1.405&    -1.04  &  8.5&  4.5&  3.408&    0.352\\
6.5&  5  &  0.277&    -1.937&   7.5&  5  &  1.336&    -0.987 &  8.5&  5  &  3.375&    0.379\\
6.5&  5.5&  0.226&    -1.794&   7.5&  5.5&  1.262&    -0.933 &  8.5&  5.5&  3.339&   0.406\\
6.5&  6  &  0.187&    -1.646&   7.5&  6  &  1.183&    -0.876 &  8.5&  6  &  3.299&    0.436\\
6.5&  6.5&  0.157&    -1.497&   7.5&  6.5&  1.101&    -0.816 &  8.5&  6.5&  3.256&    0.467\\
6.5&  7  &  0.134&    -1.35&    7.5&  7  &  1.018&    -0.752 &  8.5&  7  &  3.209&   0.5\\
6.5&  7.5&  0.115&    -1.204&   7.5&  7.5&  0.937&    -0.683 &  8.5&  7.5&  3.159&    0.536\\
6.5&  8  &  0.1  &    -1.061&   7.5&  8  &  0.858&    -0.608 &  8.5&  8  &  3.106&    0.575\\
6.5&  8.5&  0.088&    -0.92 &   7.5&  8.5&  0.783&    -0.527 &  8.5&  8.5&  3.05 &    0.616\\
6.5&  9  &  0.078&    -0.781&   7.5&  9  &  0.714&    -0.441 &  8.5&  9  &  2.99 &    0.661\\
6.5&  9.5&  0.069&    -0.645&   7.5&  9.5&  0.651&    -0.35  &  8.5&  9.5&  2.928&    0.708\\
6.5&  10 &  0.062&    -0.51 &   7.5&  10 &  0.593&    -0.254 &  8.5&  10 &  2.863&    0.759\\
6.5&  15 &  0.027&    0.692 &   7.5&  15 &  0.268&    0.77   &  8.5&  15 &  2.109&    1.349\\
6.5&  20 &  0.015&    1.605 &   7.5&  20 &  0.15 &    1.638  &  8.5&  20 &  1.402&    1.94\\
6.5&  25 &  0.01 &    2.287 &   7.5&  25 &  0.095&    2.303 &   8.5&  25 &  0.936&    2.467\\
6.5&  30 &  0.007&    2.814 &   7.5&  30  & 0.066&    2.824  &  8.5&  30 &  0.656&    2.92\\
\end{tabular}
\end{table}
\vspace{-\baselineskip} 
\noindent

\begin{table}[H]
	\centering
\begin{tabular}{cccc|cccc|cccc} \\ \hline
ADen & $T_{9}$ & EFermi & E-Cap  & ADen & $T_{9}$ & EFermi & E-Cap & ADen & $T_{9}$ & EFermi & E-Cap  \\ \hline
9&   0.5& 5.179&   0.946&   9.5& 8.5& 7.351&   2.179&   10.5&    4.5& 16.28&   3.907\\
9&   1  & 5.176&   0.969&   9.5& 9  & 7.323&   2.201&   10.5&    5  & 16.273&  3.911\\
9&   1.5& 5.17 &   1.012&   9.5& 9.5& 7.293&   2.225&   10.5&    5.5& 16.265&  3.915\\
9&   2  & 5.162&   1.051&   9.5& 10 & 7.261&   2.252&   10.5&    6  & 16.256&  3.919\\
9&   2.5& 5.151&   1.082&   9.5& 15 & 6.86 &   2.627&   10.5&    6.5& 16.247&  3.923\\
9&   3  & 5.138&   1.107&   9.5& 20 & 6.307&   3.018&   10.5&    7  & 16.237&  3.928\\
9&   3.5& 5.122&   1.128&   9.5& 25 & 5.624&   3.33 &   10.5&    7.5& 16.226&  3.934\\
9&   4  & 5.105&   1.148&   9.5& 30 & 4.859&   3.581&   10.5&    8  & 16.214&  3.94\\
9&   4.5& 5.085&   1.167&   10 & 0.5& 11.118&  2.874&   10.5&    8.5& 16.202& 3.948\\
9&   5  & 5.062&   1.185&   10 & 1  & 11.116&  2.889&   10.5&    9  & 16.189&  3.958\\
9&   5.5& 5.037&   1.204&   10 & 1.5& 11.113&  2.916&   10.5&    9.5& 16.175&  3.969\\
9&   6  & 5.01 &   1.224&   10 & 2  & 11.11 &  2.94 &   10.5&    10 & 16.16 &  3.983\\
9&   6.5& 4.98 &   1.245&   10 & 2.5& 11.105&  2.958&   10.5&    15  &15.973& 4.215\\
9&   7  & 4.948&   1.268&   10 & 3  & 11.099&  2.971&   10.5&    20 & 15.711&  4.495\\
9&   7.5& 4.914&   1.293&   10 & 3.5& 11.091&  2.981&   10.5&    25 & 15.375&  4.721\\
9&   8  & 4.878&   1.32 &   10 & 4  & 11.083&  2.99 &   10.5&    30 & 14.965&  4.892\\
9&   8.5& 4.839&   1.35 &   10 & 4.5& 11.074&  2.997&   11&  0.5& 23.934&  4.685\\
9&   9  & 4.797&   1.383&   10 & 5  & 11.063&  3.004&   11&  1  & 23.933&  4.697\\
9&   9.5& 4.754&   1.419&   10 & 5.5& 11.052&  3.01 &   11&  1.5& 23.932&  4.719\\
9&   10 & 4.708&   1.457&   10 & 6  & 11.039&  3.016&   11&  2  & 23.93 &  4.738\\
9&   15 & 4.131&   1.936&   10 & 6.5& 11.025&  3.023&   11&  2.5& 23.928&  4.752\\
9&   20 & 3.39 &   2.4  &   10 & 7  & 11.011&  3.031&   11&  3  & 23.925&  4.762\\
9&   25 & 2.621&   2.788&   10 & 7.5& 10.995&  3.04 &   11&  3.5& 23.922&  4.769\\
9&   30 & 1.973&   3.132&   10 & 8  & 10.978&  3.05 &   11&  4  & 23.918&  4.774\\
9.5& 0.5& 7.583&   1.906&   10 & 8.5& 10.959&  3.062&   11&  4.5& 23.913&  4.778\\
9.5& 1  & 7.581&   1.924&   10 & 9  & 10.94 &  3.075&   11&  5  & 23.908& 4.782\\
9.5& 1.5& 7.577&   1.958&   10 & 9.5& 10.92 &  3.091&   11&  5.5& 23.903&  4.785\\
9.5& 2  & 7.571&   1.988&   10 & 10 & 10.898&  3.11 &   11&  6  & 23.897&  4.788\\
9.5& 2.5& 7.564&   2.012&   10 & 15 & 10.624&  3.398&   11&  6.5& 23.891&  4.791\\
9.5& 3  & 7.555&   2.03 &   10 & 20 & 10.241&  3.726&   11&  7  & 23.884&  4.794\\
9.5& 3.5& 7.545&   2.044&   10 & 25 & 9.751 &  3.989&   11&  7.5& 23.877&  4.798\\
9.5& 4  & 7.532&   2.057&   10 & 30 & 9.163 &  4.194&   11&  8  & 23.869&  4.803\\
9.5& 4.5& 7.519&   2.068&   10.5&    0.5& 16.31&   3.804&   11&  8.5& 23.86&   4.809\\
9.5& 5  & 7.503&   2.08 &   10.5&    1  & 16.309&  3.816&  11&  9  & 23.851&  4.817\\
9.5& 5.5& 7.486&   2.091&   10.5&    1.5& 16.307&  3.84 &   11&  9.5& 23.842&  4.826\\
9.5& 6  & 7.468&   2.102&   10.5&    2  & 16.304&  3.861&   11& 10  &23.832 & 4.837\\
9.5& 6.5& 7.448&   2.115&   10.5&    2.5& 16.301&  3.876&   11&  15&  23.704&  5.039\\
9.5& 7  & 7.426&   2.128&   10.5&    3  & 16.297&  3.887&   11&  20&  23.526&  5.289\\
9.5& 7.5& 7.403&   2.143&   10.5&    3.5& 16.292&  3.895&   11&  25&  23.296&  5.49\\
9.5& 8  & 7.378&   2.16 &   10.5&    4   &16.286&  3.902 &  11&  30&  23.016&  5.638\\ \hline

\end{tabular}
\end{table}

\begin{figure}[htbp]
	\centering
	\includegraphics[width=0.8\textwidth]{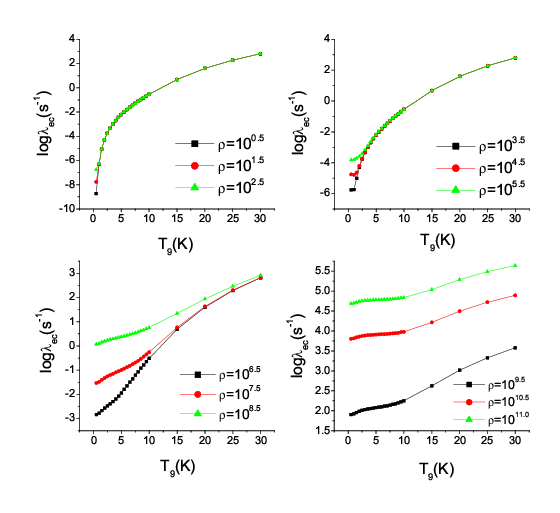}
	\caption{Electron capture rates on $^{55}$Co, as a function of
		temperature, for different selected densities. Densities are in
		units of $g\,cm^{-3}$. Temperatures are measured in $10^{9}$ K and
		log$\lambda_{ec}$ represents the log of electron capture rates in
		units of $sec^{-1}$.}
	\label{figure1}
\end{figure}
\begin{figure}[htbp]
	\centering
	\includegraphics[width=0.8\textwidth]{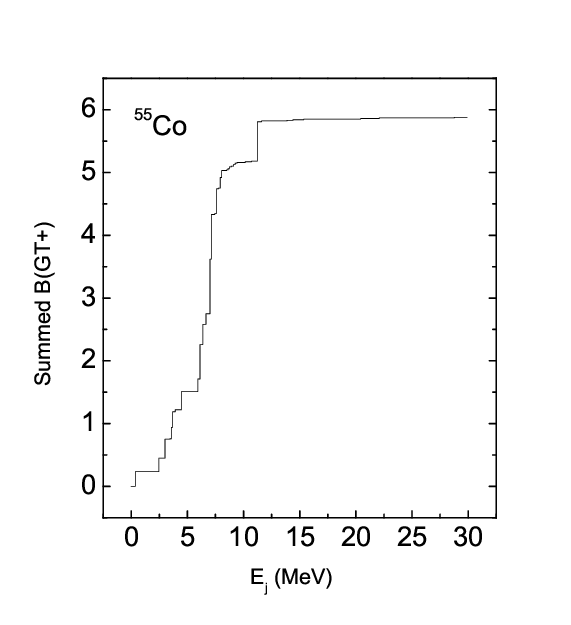}
	\caption{Cumulative sum of the B(GT) values. The energy scale refers to daughter excitation energies in $^{55}$Fe.}
	\label{figure2}
\end{figure}

\end{document}